\documentstyle[prb,aps,epsf,floats,twocolumn]{revtex}

\begin{document}
\draft

\author{Akihisa Koga, Kouichi Okunishi and Norio Kawakami}
\address{Department of Applied Physics, 
Osaka University, Suita, Osaka 565-0871, Japan}
\title{First-order quantum phase transition 
in the orthogonal-dimer spin chain}

\date{\today}
\sloppy

\wideabs{
\maketitle

\begin{abstract}

We investigate the low-energy properties of the orthogonal-dimer 
spin chain characterized by a frustrated dimer-plaquette structure.
When the competing antiferromagnetic couplings are varied, 
the first-order quantum phase transition occurs between the dimer 
and the plaquette phases, which is accompanied by nontrivial features 
due to frustration: besides the discontinuity in the lowest
excitation gap at the transition point, a sharp level-crossing 
occurs for the spectrum in the plaquette phase. We further reveal 
that the plateau in the magnetization curve at 1/4 of the full 
moment dramatically changes its character in the vicinity of the 
critical point. It is argued that the first-order phase transition
in this system captures some essential properties found in the 
two-dimensional orthogonal-dimer model proposed 
for $\rm SrCu_2(BO_3)_2$.

\end{abstract}

\pacs{PACS numbers: 75.10Jm, 75.40Cx}  
}

\narrowtext

\section{Introduction}
Recent extensive experimental and theoretical investigations on 
low-dimensional spin systems with frustration have been providing 
a variety of interesting topics. Among others, the spin-gap compound 
found recently by Kageyama et al.\cite{Kageyama}, $\rm SrCu_2(BO_3)_2$, 
 exhibits a number of nontrivial 
properties due to strong frustration. 
\cite{Kageyama,Onizuka,Nojiri,Room,Lemmens}
In particular, the discovery 
of the magnetization plateaus at 1/3, 1/4 and 1/8 of the 
full moment\cite{Kageyama,Onizuka} has been stimulating further 
intensive studies. 
\cite{Miyahara,Momoi,Fukumoto}
The remarkable point claimed by Miyahara and 
Ueda\cite{Miyahara} is that this frustrated system is a prototypical 
example of the two-dimensional (2D) version of 
{\it the orthogonal-dimer model}, \cite{Ueda}
whose unique structure gives rise to various unusual properties. 
\cite{Mila,Weihong,Muller,Vojta}
This 2D model is known to be equivalent to the frustrated Shastry-Sutherland 
model on a square lattice with some diagonal bonds.\cite{Shastry} 
It has been recently shown that there exists a novel first-order 
quantum phase transition between the dimer and plaquette phases 
in the 2D orthogonal-dimer model, which is accompanied by the jump 
and cusp singularities in the spin gap 
near the transition point.\cite{Koga}  What is most interesting is 
that the compound $\rm SrCu_2(BO_3)_2$ may be located in the vicinity 
of the first-order transition point.\cite{Kageyama,Miyahara}

As seen from the above studies, characteristic properties in the 2D 
system around the first-order transition point 
are certainly caused by the strong frustration common to
this class of the orthogonal-dimer spin systems.  Therefore,
to clarify the essential properties of the system, further 
systematic studies on the quantum phase transition are highly desired.  
The 1D version of the orthogonal-dimer model\cite{Kato,Ivanov,Koga1}
 may be the most appropriate
system for this purpose, because it is the simplest model which possesses
the frustration effect due to the dimer-plaquette structure.
\begin{figure}[htb]
\epsfxsize=8cm
\centerline{\epsfbox{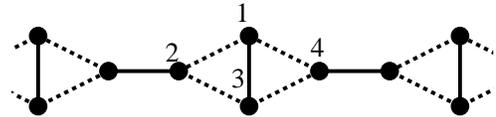}} 
\vspace{0.1cm}
\caption{The orthogonal-dimer chain. 
The solid and dashed lines indicate the antiferromagnetic exchange 
couplings $J$ and $J'$, respectively.}
\label{fig:plachn}
\end{figure}
Motivated by these hot  topics,
in this paper, we study the first-order transition in 
the 1D orthogonal-dimer 
spin chain in detail.\cite{Kato,Ivanov,Koga1} 
The Hamiltonian we shall deal with is,
\begin{eqnarray}
{\cal H}=J\sum_{(i,j)}{\bf S}_{i}\cdot{\bf S}_{j}+
J'\sum_{<i,j>}{\bf S}_{i}\cdot{\bf S}_{j}-H\sum_{i}S_{i}^{z}, 
\label{hamiltonian}
\end{eqnarray}
where ${\bf S}_{i}$ is the $s=1/2$ spin operator at the $i$-th site, and 
the indices $(i,j)$ and $<i,j>$ represent the summation over intra- and 
inter-dimer pairs, respectively (see Fig. \ref{fig:plachn}).
The magnetic field is denoted as $H$ for which we set
 $g\mu_{\rm B}=1$ for convenience.  Both of
the exchange couplings  $J $ and $J'$ are assumed to be antiferromagnetic. 
We shall use the normalized parameters $j=J'/J$ and $h=H/J$ in the 
following discussions.

It was previously shown by Ivanov and Richter\cite{Ivanov}
 that the above orthogonal-dimer chain undergoes the first-order 
quantum phase transition between the dimer phase (for small $j$) and 
the plaquette phase (for large $j$) as the coupling ratio $j$ is varied.
However, the unique properties inherent in this frustrated system
have not been discussed in detail yet, especially around the critical point.  
In what follows, by means of the exact diagonalization (ED) and series 
expansion methods,
we demonstrate that the excitation gap in this system exhibits
 nontrivial behaviors such as the jump and cusp singularities around the 
transition point, reflecting the dimer-plaquette dual properties
characteristic of  this frustrated system.  We also point out that 
such properties are not specific to the 1D system but also 
common to the 2D orthogonal-dimer system.
By computing the magnetization curve by means of
the ED  together with the density-matrix renormalization group 
(DMRG),\cite{White,pwfrg} 
we further reveal that the formation of the magnetic plateaus is 
dramatically affected by the dimer-plaquette dual properties, and the 
resulting magnetic phase diagram has  a rich structure.

This paper is organized as follows. 
In Sec. \ref{sec:QPT}, we investigate how the characteristic properties 
inherent in the orthogonal-dimer chain emerge in the low-energy 
excitations around the first-order transition point
by means of the ED, the DMRG\cite{White,pwfrg} and the series expansion 
method.\cite{Singh} In Sec. \ref{sec:4}, by  calculating the 
magnetization curve by the ED and the DMRG, we show that the 
plateau formation in the magnetization changes its character 
around the transition point, reflecting the strong frustration effects.
The last section is devoted to  summary and discussions.

\section{zero-field properties}\label{sec:QPT}
In this section, we investigate the quantum phase transition 
in the orthogonal-dimer chain in the absence of
 the magnetic field,\cite{Ivanov}  by exploiting the ED  
and the series expansion methods.\cite{Singh}

\subsection{Ground state}

\begin{figure}[htb]
\epsfxsize=8cm
\centerline{\epsfbox{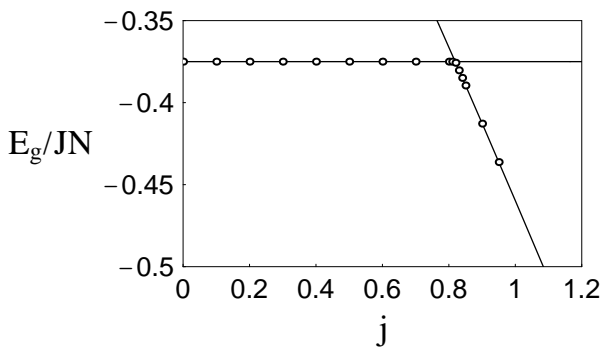}} 
\vspace{0.1cm}
\caption{The ground state energy per site for the orthogonal-dimer chain 
as a function of $j(=J'/J)$. 
The flat line ($E_{\rm g}/JN=-3/8$) indicates 
the exact ground state energy of the dimer phase.
The solid line is the result obtained by the plaquette expansion.
We also show the results for the finite chain of $N=24$ as open circles.}
\label{fig:eg}
\end{figure}

We start with the ground state properties in 
the dimer phase. When $j=0$, the system is reduced to an assembly of 
the decoupled dimers denoted by the solid line 
in Fig. \ref{fig:plachn}, for which the product of 
independent dimer-singlets gives the ground state.
The remarkable point for the orthogonal-dimer chain
\cite{Miyahara,Ivanov} is that 
this simple dimer-singlet state is always an exact eigenstate of 
the Hamiltonian (\ref{hamiltonian}) with the energy
$E_{\rm g}/JN=-3/8$ in the entire range of $j$ (see Fig. \ref{fig:eg}),
where $N$ is the number of total sites.
Accordingly, the dimer-singlet state should be the exact ground state up to 
a certain critical value of $j$.

In the opposite limit of large $j$,  
the ground state is given by the disordered singlet state which 
is adiabatically connected to the isolated plaquette-singlets 
denoted by the broken line in Fig. \ref{fig:plachn}. Therefore,
starting from the isolated plaquette singlets for $j=\infty$, we can
 evaluate the ground state energy $E_{\rm g}$ 
for the plaquette phase with finite $j$ by means of the series expansion.
Performing the plaquette expansion up to the eleventh order in $j^{-1}$ 
combined with the first-order inhomogeneous differential method,\cite{Pade} 
we have obtained the ground state energy $E_{\rm g}$ rather
precisely for the plaquette phase.
The result is shown as the solid line in Fig. \ref{fig:eg}.
It is seen  that the ground state energy of the plaquette state 
coincides with that of the dimer state 
at a certain value of $j$, at which  the first-order quantum 
phase transition occurs.
The critical value is estimated as $j_{c}=0.81900$, 
which is further confirmed to be accurate up to the above figure
 by the DMRG calculation for the infinite chain.
We note here that the present results for the ground state are 
consistent with those obtained by Ivanov and Richter.\cite{Ivanov}

\subsection{Spin excitations}

\begin{figure}[htb]
\epsfxsize=8cm
\centerline{\epsfbox{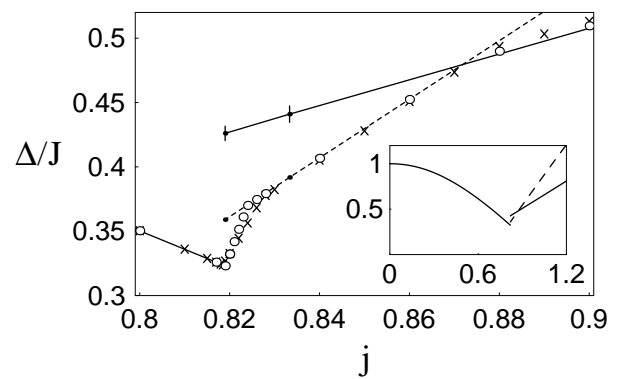}} 
\vspace{0.1cm}
\caption{The spin gap $\Delta/J$ for the orthogonal-dimer chain 
as a function of $j(=J'/J)$.
The crosses and open circles indicate the results obtained by the ED method 
for the finite chain of $N=16$ and $24$, respectively. See the 
text  about the solid and broken lines.}
\label{fig:gap2}
\end{figure}

Let us move to the spin excitation spectrum.
In the dimer phase $(0<j<j_c)$, 
we can construct a low-energy triplet excitation by substituting a local 
triplet for one of singlet-dimers forming the ground state.
The remarkable point is that the hopping of 
this local triplet across the singlet-dimer such as
 the $1$-$3$ dimer in Fig. 1 is forbidden 
by its characteristic crystal structure.\cite{Miyahara}
This implies that an excited triplet state is completely localized  inside
the finite strip $(N=6)$, and has no dispersion for its spectrum.
Thus we can exactly estimate the spin gap from the ED calculation for 
the cluster of $N=6$.
In Fig. \ref{fig:gap2}, the ED results are shown as the solid line in the
dimer phase.

In the plaquette phase $(j>j_c)$, it is naively expected that
the low-energy excitation is described in terms of  a 
triplet excitation on the isolated plaquette.
However, the frustrating diagonal interaction in each plaquette 
should considerably affect the excitation spectrum.
To see this, it is instructive to examine the energy-level 
structure of the isolated plaquette with the diagonal 
bond (see Fig. \ref{fig:plachn}), whose Hamiltonian reads 
\begin{eqnarray}
{\cal H}_{\rm plaquette}=J {\bf s}_1\cdot{\bf s}_3+
J'\left({\bf s}_1+{\bf s}_3\right)\cdot\left({\bf s}_2+{\bf s}_4\right). 
\label{iplaq}
\end{eqnarray}
We list the energy eigenvalues of the plaquette in Table \ref{TableI}, 
where $S_{13}$ ($S_{24}$) is a quantum number of the spin 
${\bf s}_1+{\bf s}_3$ $({\bf s}_2+{\bf s}_4)$ and
$S$ is that for the total spin.
\begin{table}[htb]
\caption{Eigenstates of an isolated plaquette.}
\begin{tabular}{lccc|cccc}
         & \multicolumn{2}{c}{$D$-sector}   &&
\multicolumn{4}{c}{$P$-sector} 
\\ \hline
$S_{13}$ & 0 & 0 && 1 & \multicolumn{3}{c} 1 \\
$S_{24}$ & 0 & 1 && 0 & \multicolumn{3}{c} 1 \\ \cline{6-8}
$S     $ & 0 & 1 && 1 & 0 & 1 & 2 \\ \hline
$E/J$ & $-\frac{3}{4}$ & $-\frac{3}{4}$&& $\frac{1}{4}$ 
         & $-2j+\frac{1}{4}$ & $-j+\frac{1}{4}$ &
           $j+\frac{1}{4}$ \\
\end{tabular}\label{TableI}
\end{table}
It is seen in this table that the eigenstates of the isolated 
plaquette are classified into two sectors: 
the $D$-sector for $S_{13}=0$ and the $P$-sector for $S_{13}=1$. 
For $j>1/2$, the ground state is always singlet
in the $P$-sector, whereas both of two separated sectors appear
for the excitations when the coupling ratio $j$ is varied.
When $1/2<j<1$, the lowest excitation is given by the $S_{24}$-singlet 
and $S_{24}$-triplet in the $D$-sector, 
which are four-fold degenerate with the excitation energy
 $\Delta E/J = 2j-1 $.  In the regime $j>1$, 
the first excitation is a $P$-sector triplet with 
the excitation energy  $\Delta E/J=j$.  Note that
as far as the singlet-triplet excitations in the $P$-sector are concerned,
there are no contributions from the $D$-sector. 
This simple fact allows us to replace $S_{13}$ in the plaquette with 
the $s=1$ spin, which will be used below to perform the mixed-spin cluster 
expansion for the $P$-sector.

Keeping the above properties in mind, we now turn back to the 
orthogonal-dimer chain in the thermodynamic limit, 
by taking into account the interaction among plaquettes.
Note that excitations in the $D$-sector are still 
completely separated from those in the $P$-sector, according to the 
constraint specific to the orthogonal-dimer structure. 
This remarkable fact enables us to study the excitation
spectrum for the $D$-sector and the $P$-sector separately.
Namely, we perform the plaquette expansion up to the eleventh order 
in $j^{-1}$ for the $D$-sector, whereas for the $P$-sector
we exploit  the mixed-spin cluster expansion\cite{exkoga} up to 
the seventh order in $j^{-1}$ 
with a starting spin configuration $1/2\circ 1 \circ 1/2$.\cite{exkoga}
Applying the first-order inhomogeneous differential method\cite{Pade} 
to the obtained series, we then deduce the lowest excitations
both  for the $D$-sector and the $P$-sector, as shown 
in Fig. \ref{fig:gap2}. 
In this figure, the dashed line in the plaquette phase 
represents the energy for
four-fold degenerate excitations in the $D$-sector whereas 
the solid line is for a triplet excitation in the $P$-sector.
It is seen that the energy of two kinds of excitations intersects
 each other
at $j'_c=0.872$, giving rise to  a cusp singularity in the spin gap 
as a function of $j$.
In the region $j>j'_c$, both of the ground state and the 
lowest excited state 
belong to the $P$-sector  on each plaquette, 
as is the case for the simple plaquette chain.\cite{Kato}
On the other hand, when $j_c<j<j'_c$, 
the lowest excitation is described by four-fold degenerate level 
in the $D$-sector although the ground state still belongs 
to the plaquette phase ($P$-sector).
The present plaquette expansion further uncovers that the 
wave function of the localized $D$-sector excitation with no dispersion
is spatially extended over a number of
sites, which shows  sharp contrast to the dimer phase
where the wave function of the triplet excitation is
completely localized at a given site.
This may imply that the four-fold degenerate excitations characterized 
by the $D$-sector possess the intermediate properties 
between those typical for  the dimer phase and the plaquette phase.
As $j$ is further decreased, we encounter the first-order quantum 
phase transition,  at which the ground-energy of the two phases 
coincide with each other, while the first excited states have still
different energies.  Therefore,
 the spin gap at the transition point jumps from 
$\Delta^+(j_{c})=0.3590(2)$ for the plaquette phase 
to $\Delta^-(j_{c})=0.32309$ for the dimer phase.

In order to complement the series-expansion results, we also compute 
the lowest triplet excitation by the ED method for the finite 
chain ($N=16, 24$) with the periodic boundary condition.
The obtained spin gap $\Delta$ is shown as the crosses and the 
open circles in Fig. \ref{fig:gap2}.
It is seen that they are in fairly good agreement with the results obtained by
the series expansion method, except for the vicinity of the first-order 
transition point $j_c$, where the finite-size effect in the ED still remains.
Here it should be noticed that the present results 
provide much more detailed information for spin excitations
than those obtained by Ivanov and Richter, where the spin 
gap as a function of $j$ has only a cusp-like singularity at the 
critical point.\cite{Ivanov}
We also note that the ladder system with a similar orthogonal-dimer 
structure has the jump and the cusp singularities 
in the spin gap.\cite{Gellad}

\section{Plateaus in the magnetization curve}\label{sec:4}
In this section, we study the magnetic properties  of 
the orthogonal-dimer chain. 
 By using the ED and the DMRG\cite{White,pwfrg} methods, 
we calculate the magnetization curve 
as a function of  the magnetic field $h(=H/J)$, and  
show that  their characteristic behavior
stems from the dimer-plaquette dual properties 
found for the excitations in the previous section.
To clearly understand the magnetization curve in the low-field region,
we shall use the description based on hard-core bosons, 
in which low-energy triplet excitations are regarded as 
hard-core bosons.

\subsection{Dimer phase}

\begin{figure}[htb]
\epsfxsize=8cm
\centerline{\epsfbox{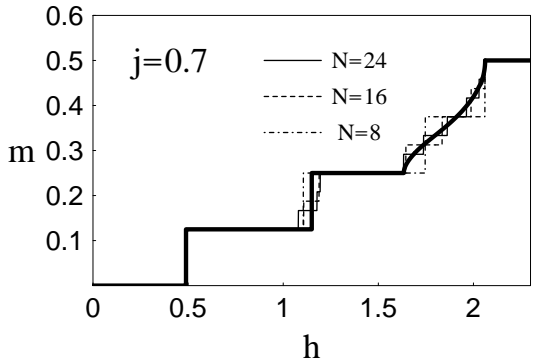}} 
\vspace{0.1cm}
\caption{The magnetization $m(=M/N)$ as a function of $h$ 
in the dimer phase $(j=0.7)$.
Three thin lines represent the magnetization curves for 
the finite chain of $N=8, 16$ and $24$.
The bold line is the result obtained by the DMRG method 
for the infinite chain.}
\label{fig:7}
\end{figure}

Let us start with the dimer phase.
In Fig. \ref{fig:7}, we show the magnetization curve for $j=0.7(<j_c)$ 
obtained by the ED method for the finite chain of $N=8, 16$ and $24$.
We also show the results obtained by the DMRG method.
In this figure, the plateaus appear in the magnetization curve
at 1/4 and 1/2 of the full moment ($m=1/2$) clearly.
 The mechanism of the $1/2$-plateau formation is simply understood
according to  the unit-cell structure of the chain.
As will be shown  below, the $1/2$-plateau appears  in 
the entire range of $j$.
On the other hand, the $1/4$-plateau may be realized by the commensurate 
state in which each triplet excitation is  aligned periodically
 at every finite-strip of $N=8$, 
reflecting the completely-localized nature of 
the triplet excitation discussed in 
the previous section.
Since the localized triplet does not correlate with those in the 
neighbor strips, the macroscopic number of the triplets become 
degenerate at the lower-critical field $h_c$ (= the spin gap $\Delta$), 
where the magnetic first-order transition occurs.
In Fig. \ref{fig:7}, 
we can indeed see that the magnetization curve jumps from $m=0$ to $m=1/8$ 
(1/4-plateau) at $h_c$ without any finite-size correction.
Increasing the magnetic field,
the other first-order transition to the 1/2-plateau state occurs.
It is seen in Fig. \ref{fig:7} that there still remains the large-finite size 
effect near this critical point, since the triplet excitations correlate with 
each other in contrast to the transition discussed above.

We here note that several different phases observed 
in the magnetization process
are specified by the spatial arrangement of the spin quantum number, 
$S_{13}$, for the diagonal bond in each plaquette (see Table \ref{TableI}).
In the dimer phase, 
the $D$-sector ($S_{13}$=0) is realized in each plaquette and
in the 1/4-plateau state, the $D$- and the $P$-sectors 
are crystallized alternately.
On the other hand, in the region of $m>1/4$, 
each diagonal bond in the plaquette forms $S_{13}=1$
(the $P$-sector),
and hence the frustration does not play any role for the magnetization process
in this region.
Therefore by calculating the magnetization for 
the unfrustrated plaquette chain,
we can simply reproduce that for the orthogonal-dimer chain in this region.

\subsection{Plaquette phase}

We recall that there are two sort of 
excitations characterized by the $P$- and the $D$-sector in the
plaquette phase, which 
may provide a variety of the magnetization process.
More precisely, the hard-core boson description suggests 
 three possible behaviors for the magnetization curves:
\begin{itemize}
\item
 case I: $D$-triplet excitations lie energetically lower than 
$P$-triplet excitations.
The magnetization curve has a structure similar to that in the dimer-phase.

\item
 case II: $D$-triplet excitations lie slightly above the bottom 
of the $P$-triplet dispersion curve (see Fig. \ref{fig:dis94}).
Thus both of the features specific to the  $D$-triplet and the $P$-triplet 
excitations appear in the magnetization curve.

\item
 case III: $D$-triplet excitations lie sufficiently higher above the
spin gap for $P$-triplet excitations.
The magnetization curve may be characterized solely 
by the $P$-triplet excitation.

\end{itemize}
Since the above classification is simply based on the hard-core
boson description, 
the interaction among  bosons becomes more important
when the number of bosons (i.e. the strength of the external field) 
is increased, which requires more proper discussions beyond 
the hard-core boson description. 
In what follows, based on the numerical results,
 we confirm that the magnetization curves 
are indeed classified into the above three categories.

\begin{figure}[htb]
\epsfxsize=8cm
\centerline{\epsfbox{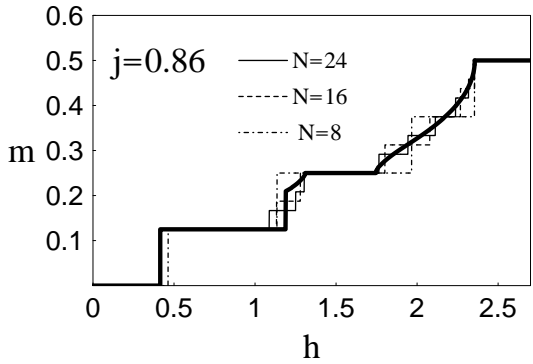}} 
\vspace{0.1cm}
\caption{The magnetization $m(=M/N)$ as a function of $h$ 
in the plaquette phase $(j=0.86)$.
Three thin lines represent the magnetization curves for 
the finite chain of $N=8, 16$ and $24$.
The bold line is the result obtained by the DMRG for the infinite chain.}
\label{fig:86}
\end{figure}

\begin{figure}[htb]
\epsfxsize=8cm
\centerline{\epsfbox{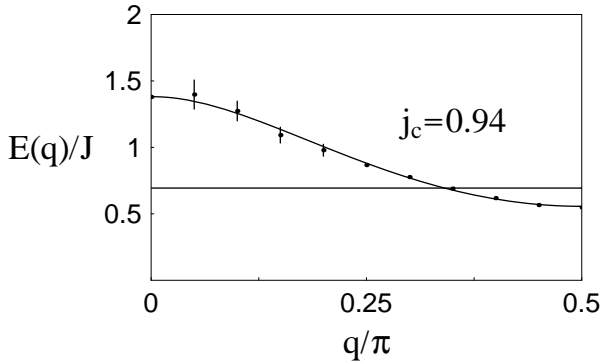}} 
\vspace{0.1cm}
\caption{The dispersion relation for  $j=0.94$. 
The flat line and the solid curve
represent the four-fold degenerate $D$-triplet excitations
and the $P$-triplet  excitations, 
which are obtained by the mixed-spin cluster expansion and the plaquette
expansion, respectively.
}
\label{fig:dis94}
\end{figure}

We start with the analysis of the case I.
We show the magnetization process for $j=0.86$ in Fig. \ref{fig:86}, 
where we can see the $1/4$- and $1/2$-plateaus clearly.
In particular the magnetization jumps from zero to the $1/4$ plateau at the 
critical field $h_c$ with a small finite-size correction.
In the plaquette phase, the 1/4-plateau is generated by 
the commensurate state in which the plaquette singlet and triplet 
in the $D$-sector are crystallized alternately.
The characteristic point in the 1/4-plateau state is that 
$D$-triplet excitations interact with each other when they are 
condensed to form the $1/4$-plateau, in contrast to those in the 
dimer phase where each triplet excitation is localized completely. 
We can indeed see this interaction effect in the ED spectrum, 
where the triplets gain the condensation energy,
by which the critical field $h_c$ shifts to a slightly lower 
field than the spin-gap value. We  note  here that 
since the mechanism of the 1/4-plateau formation (first-order
transition) at $h_c$ is essentially the same in the dimer phase and
in the plaquette phase, so that the critical field 
$h_c$  does not show the discontinuity at $j=j_c$ as a function of $j$
(see Fig.\ref{fig:phase}),
being contrasted to the discontinuity observed 
in the spin gap at $j=j_c$.
Increasing the field, the magnetic first-order transition takes place to 
the state in which $S_{13}=1$ is formed in each plaquette.
Beyond this critical field, the magnetization traces the curve 
which is the same as that for unfrustrated plaquette chain.

\begin{figure}[htb]
\epsfxsize=8cm
\centerline{\epsfbox{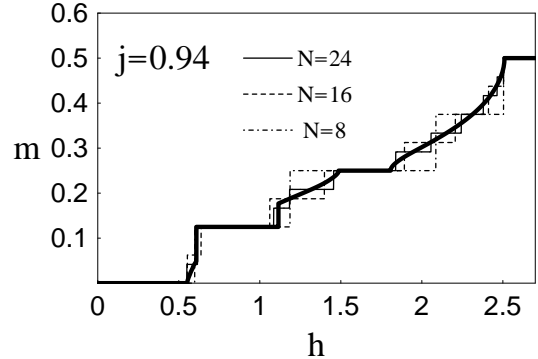}} 
\vspace{0.1cm}
\caption{The magnetization $m(=M/N)$ as a function of $h$ 
in the plaquette phase $(j=0.94)$.
Three thin lines represent the magnetization curves for 
the finite chain of $N=8, 16$ and $24$.
The bold line is the result obtained by the DMRG for the infinite chain.}
\label{fig:94}
\end{figure}

In the case II, the $D$-triplet excitation level lies slightly above 
the bottom of the $P$-triplet dispersion curve, as shown 
in Fig.\ref{fig:dis94}.
The coexistence of two-kinds of distinct excitations influences 
the magnetization curve considerably, as seen in Fig. \ref{fig:94}.
As $h$ increases beyond the critical field $(h_c=\Delta)$, 
the magnetization should develop with $m\sim (h-h_c)^{1/2}$, 
since it is dominated by the $P$-triplet excitation whose 
dispersion relation is quadratic near the bottom $(q=\pi/2)$.
In fact, we can see a staircase structure below 
the $1/4$-plateau for the ED results in Fig. \ref{fig:94}, implying that the 
jump is now changed to the continuous increase in the magnetization.
With slightly increasing $h$, however, the magnetization may stop 
to increase continuously and jump to the $1/4$-plateau.
Beyond the 1/4-plateau, we again encounter the first-order phase transition,
which is accompanied by the jump in $m$.
This jump singularity is followed by the continuous increase in $m$,
as already discussed in the case I.

In the case III,  the excitation energy of the $D$-triplet 
is pushed up in higher energy region of the dispersion curve.
Thus, before the Zeeman energy lowers the $D$-triplet level down to zero,
a number of bosons are accommodated in the $P$-triplet excited levels, 
and then the interaction effect between bosons becomes 
too significant to use the hard-core boson description.
In this parameter region, we find that  the $1/4$-plateau 
completely disappears  and the magnetization curve 
becomes smooth up to the $1/2$-plateau. 
\begin{figure}[htb]
\epsfxsize=8cm
\centerline{\epsfbox{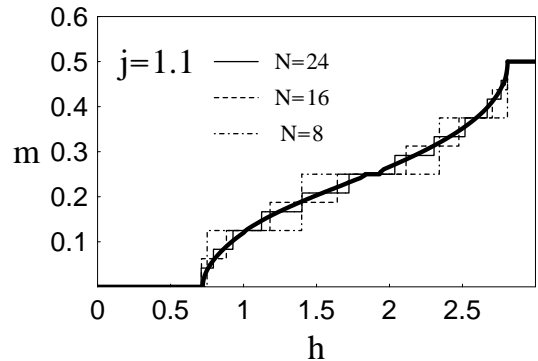}} 
\vspace{0.1cm}
\caption{The magnetization $m(=M/N)$ as a function of $h$ 
in the plaquette phase $(j=1.1)$.
Three thin lines represent the magnetization curves for 
the finite chain $N=8, 16$ and $24$.
The bold line is the result obtained by the DMRG for the infinite chain.
Note that the 1/2-plateau state still exists near $h=1.9$.
}
\label{fig:11}
\end{figure}

Consequently, we end up with  the magnetic phase diagram on the $(j-h)$ plane 
as shown in Fig. \ref{fig:phase}. 
\begin{figure}[htb]
\epsfxsize=8cm
\centerline{\epsfbox{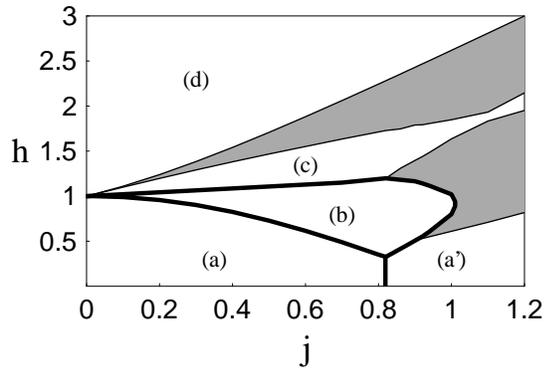}} 
\vspace{0.1cm}
\caption{The ground-state phase diagram on the $j$ versus $h$ plane.
The phase boundaries are determined by the ED method for the finite chain 
of $N=24$ and the DMRG method for the infinite chain. 
}
\label{fig:phase}
\end{figure}
In this figure, (a) and (a') indicate 
the dimer-singlet  and plaquette-singlet phase, respectively.
The region (b) exhibits the plateau in the magnetization curve at 1/4 
of the full moment.
We can see that the 1/4-plateau phase with the dimer-singlet structure 
exists not only in the dimer phase $(j<j_c)$ but also 
in a part of the plaquette phase.
In the region (c), the $1/2$-plateau appears, 
which is stable in the whole parameter region of $j$.
All spins are polarized in (d).  
The bold (solid) line shows the phase boundary of 
the first- (second-) order transition.
It is now seen that the magnetic 
phase diagram possesses quite a rich structure, which is 
indeed caused by the dimer-plaquette dual properties inherent 
in this strongly frustrated spin chain.
Here we wish to mention that the ladder system with a similar orthogonal-dimer 
structure\cite{Gellad} has the jump and the cusp singularities 
in the magnetization curve.\cite{Honecker}

\section{Summary and Discussions}\label{sec:5}
We have studied the low-energy properties of the orthogonal-dimer spin chain 
with a frustrated dimer-plaquette structure, by means of the 
exact diagonalization, the density matrix renormalization group 
and the series expansion method.
When the inter-dimer couplings are varied, the first-order quantum 
phase transition  occurs at the critical 
value $j_c=0.81900$ between the dimer-singlet and plaquette-singlet phases.
In the dimer phase, the ground state is exactly given by the 
decoupled dimer-singlet state,  and triplet excitations over it are 
completely localized.
In the plaquette phase, on the other hand, we have found that the 
excited states are
characterized by the dual properties: the coexistence of plaquette-like 
and dimer-like excitations. As a result, both of the cusp and the jump 
appear in the spin gap 
as a function of $j$, which are caused by the level-crossing 
between the excited states.
We have also clarified  how the above dual properties
influence the magnetization process of the chain by analyzing 
the numerical results in terms of the hard-core boson description.
In particular, we have shown that three types of the magnetization 
curves appear around the magnetization plateau at $1/4$ of the 
full moment, according to  how two kinds 
of triplet excitations change their relative positions 
energetically.
We have thus  found that  the magnetic phase diagram  on 
the $(j-h)$ plane has a rich structure.

Although we have dealt with the one-dimensional chain in this paper,
we believe that the present study captures some essential properties
of the first-order phase transition and the resulting rich structure in
low-lying excited states even in two dimension.
For the compound $\rm SrCu_2(BO_3)_2$, it is indeed pointed out 
that the low-energy triplet excitation has quite small mobility 
reflecting the orthogonal-dimer structure.\cite{Miyahara}
Since this compound is located in the vicinity of the first-order 
phase transition point, the present results encourage us to 
interpret the plateau-formation mechanism of the 2D compound in 
terms of the dual properties originating from  the dimer and 
plaquette structures.  In particular, it is an interesting 
 problem to study how the characteristic properties in 1D are
changed into those in 2D  by adiabatically introducing  the 
 inter-chain coupling, which may provide a clue to
fully understand the physics in 2D orthogonal-dimer system.

\section*{Acknowledgement}
The work is partly supported by a Grant-in-Aid from the Ministry of 
Education, Science, Sports, and Culture. 
A. K. and K. O. are supported by the Japan Society for the Promotion 
of Science.
A part of our computational program is based 
on the TITPACK ver. 2 by H. Nishimori.
Numerical computations in this work was carried out at the Yukawa 
Institute Computer Facility.

\end{document}